\documentclass{aa}
\usepackage{graphicx}
\usepackage{color}
\usepackage{numprint}
\usepackage{txfonts}
\usepackage{xspace}

\newcommand{\system}{NGTS-14\xspace}
\newcommand{\starq}{NGTS-14A\xspace}
\newcommand{\companion}{NGTS-14B\xspace}
\newcommand{\planetq}{NGTS-14Ab\xspace}

\newcommand{\feh}{\mbox{[Fe/H]}\xspace}
\newcommand{\teff}{\mbox{$T_{\rm *, eff}$}\xspace}
\newcommand{\logg}{\mbox{$\log g_*$}\xspace}
\newcommand{\loggp}{\mbox{$\log g_{\rm p}$}\xspace}
\newcommand{\vsini}{\mbox{$v \sin i_{*}$}\xspace}
\newcommand{\kms}{\mbox{km\,s$^{-1}$}\xspace}
\newcommand{\ms}{\mbox{m\,s$^{-1}$}\xspace}

\newcommand{\mjup}{\mbox{$\mathrm{M_{\rm Jup}}$}\xspace}
\newcommand{\rjup}{\mbox{$\mathrm{R_{\rm Jup}}$}\xspace}

\newcommand{\me}{\mbox{$\mathrm{M_{\rm \oplus}}$}\xspace}

\newcommand{\mstar}{\mbox{$M_{*}$}\xspace}
\newcommand{\rstar}{\mbox{$R_{*}$}\xspace}

\newcommand{\msol}{\mbox{$\mathrm{M_\odot}$}\xspace}
\newcommand{\rsol}{\mbox{$\mathrm{R_\odot}$}\xspace}

\newcommand{\prot}{\mbox{$P_\mathrm{*,rot}$}\xspace}

\newcommand{\tlcm}{{\sc TLCM}\xspace}

\newcommand{\PERIOD}{$3.5357173 \pm 0.0000069$}
\newcommand{\EPOCH}{$7502.5545 \pm 0.0020$}
\newcommand{\DURATION}{$ 2.24\substack{+0.08 \\ -0.06}$}

\newcommand{\KAmp}{$13.2 \pm 1.7 $}
\newcommand{\GAMMA}{$30.2876 \pm 0.0013$}
\newcommand{\ARSTAR}{$10.3 \substack{+2.3 \\ -2.6}$}
\newcommand{\RPRS}{$0.0530 \substack{+0.0038 \\ -0.0033}$}

\newcommand{\IMPACT}{$0.59 \substack{+0.23 \\ -0.38}$}
\newcommand{\INCLINATION}{$86.7 \pm 1.7$}

\newcommand{\UPLUS}{$ 0.65\pm 0.19$}
\newcommand{\UMINUS}{$ 0.30\pm 0.21$}

\newcommand{\SecondaryTeff}{$1143 \pm 139$}
\newcommand{\SecondaryMass}{$0.092 \pm 0.012$}
\newcommand{\SecondaryRadius}{$0.444 \pm0.030 $}
\newcommand{\SecondaryDensity}{$1395\pm333$}
\newcommand{\SecondaryLogg}{$3.08\pm0.08$}
\newcommand{\SMA}{$0.0403 \pm 0.0071$}

\begin{document}

\title{\planetq: a Neptune-sized transiting planet in the desert}

\titlerunning{\planetq: a Neptune-sized transiting planet in the desert}

   \author{A. M. S. Smith\inst{1}\fnmsep\thanks{\email{alexis.smith@dlr.de} }
          \and J.~S.~Acton\inst{2}
          \and D.~R.~Anderson\inst{3,4}
          \and D.~J.~Armstrong\inst{3,4}
          \and D.~Bayliss \inst{3,4}
          \and C.~Belardi \inst{2}
          \and F.~Bouchy \inst{5}
          \and R.~Brahm\inst{6,7}
          \and J.~T.~Briegal\inst{8}
          \and E.~M.~Bryant\inst{3,4}
          \and M.~R.~Burleigh\inst{2}
          \and J.~Cabrera\inst{1}
          \and A.~Chaushev\inst{9}
		  \and B.~F.~Cooke\inst{3,4} 
		  \and J.~C.~Costes\inst{10}
          \and Sz.~Csizmadia\inst{1}
		  \and Ph.~Eigm\"uller\inst{1}
		  \and A.~Erikson\inst{1}
          \and S.~Gill\inst{3,4}
          \and E.~Gillen\inst{8,11,12}
          \and M.~R.~Goad\inst{2}
          \and {M.~N.~G{\"u}nther}\inst{13,14}
          \and B.~A.~Henderson\inst{2}
          \and A.~Hogan\inst{2}
          \and A.~Jord\'an\inst{6,7}
          \and M.~Lendl\inst{5}
          \and J.~McCormac\inst{3,4}
          \and M.~Moyano\inst{15}
          \and L.~D.~Nielsen \inst{5}
          \and H.~Rauer\inst{1,9,16}
          \and L.~Raynard\inst{2}
          \and R.~H.~Tilbrook\inst{2}
          \and O.~Turner\inst{5}
          \and S.~Udry\inst{5}
          \and J.~I.~Vines\inst{17}
          \and C.~A.~Watson \inst{10}
          \and R.~G.~West\inst{3,4}
          \and P.~J.~Wheatley\inst{3,4}
          }

   \institute{Institute of Planetary Research, German Aerospace Center, 
              Rutherfordstrasse 2, 12489 Berlin, Germany              
         \and School of Physics and Astronomy, University of Leicester, LE1 7RH, UK
         \and Centre for Exoplanets and Habitability, University of Warwick, Gibbet Hill Road, Coventry CV4 7AL, UK
         \and Dept.\ of Physics, University of Warwick, Gibbet Hill Road, Coventry CV4 7AL, UK
         \and Observatoire de Gen{\`e}ve, Universit{\'e} de Gen{\`e}ve, 51 Ch. des Maillettes, 1290 Sauverny, Switzerland
        \and Facultad de Ingenier\'ia y Ciencias, Universidad Adolfo Ib\'a\~{n}ez, Av.Diagonal las Torres 2640, Pe\~{n}alol\'en, Santiago, Chile
        \and Millennium Institute for Astrophysics, Santiago,Chile
         \and Astrophysics Group, Cavendish Laboratory, J.J. Thomson Avenue, Cambridge CB3 0HE, UK
         \and Center for Astronomy and Astrophysics, TU Berlin, Hardenbergstr. 36, D-10623 Berlin, Germany
         \and Astrophysics Research Centre, School of Mathematics and Physics, Queen's University Belfast, BT7 1NN Belfast, UK
        \and Astronomy Unit, Queen Mary University of London, Mile End Road, London E1 4NS, UK
         \and Winton Fellow
        \and Department of Physics, and Kavli Institute for Astrophysics and Space Research, Massachusetts Institute of Technology, 77 Mass. Ave, Cambridge, MA 02139, USA
        \and Juan Carlos Torres Fellow
         \and Instituto de Astronomía, Universidad Católica del Norte, Angamos 0610, 1270709, Antofagasta, Chile
         \and Institute of Geological Sciences, FU Berlin, Malteserstr. 74-100, D-12249 Berlin, Germany
         \and Departamento de Astronomia, Universidad de Chile, Casilla 36-D, Santiago, Chile
}

   \date{Received 19 October 2020; accepted 24 December 2020}

 
  \abstract
   {The sub-Jovian or Neptunian desert is a previously-identified region of parameter space where there is a relative dearth of intermediate-mass planets at short orbital periods. }
   {We present the discovery of a new transiting planetary system within the Neptunian desert, \starq.}
   {Transits of \planetq were discovered in photometry from the Next Generation Transit Survey (NGTS). Follow-up transit photometry was conducted from several ground-based facilities, as well as extracted from TESS full-frame images. We combine radial velocities from the HARPS spectrograph with the photometry in a global analysis to determine the system parameters.}
   {\planetq has a radius about 30 per cent larger than that of Neptune (\SecondaryRadius\,\rjup), and is
   around 70 per cent more massive than Neptune (\SecondaryMass\,\mjup). It transits the main-sequence K1
   star, \starq, with a period of 3.54 days, just far enough to have maintained at least some of its
   primordial atmosphere. We have also identified a possible long-period stellar mass companion to the
   system, \companion, and we investigate the binarity of exoplanet host stars inside and outside the
   Neptunian desert using Gaia.}
   {}
   {}


   \maketitle
%

\section{Introduction}

The first generation of wide-field transit surveys, the most prolific of which were SuperWASP \citep{Pollacco06} and HAT-Net \citep{bakos02}, unveiled the rich diversity of hot Jupiters. These planets, although intrinsically rare, have been discovered in numbers large enough to enable statistical population analyses such as those investigating inflation (e.g. \citealt{Thorngren18,Sestovic18}). Transiting hot Jupiters also remain the best-studied individual planetary systems, with characterisation observations such as atmospheric transmission spectroscopy pioneered on these objects (e.g. \citealt{Sing16}).

The Kepler \citep{Borucki10} and K2 \citep{K2} missions, and the ongoing TESS (Transiting Exoplanet Survey Satellite; \citealt{TESS}) mission subsequently revealed a large population of small (less than 2 -- 3 Earth radii), short-period planets. In between these two populations, however, lies a relatively unpopulated region of parameter space, often referred to as the sub-Jovian or Neptunian desert \citep{Szabo_Kiss_11,Mazeh16}. 

The Next Generation Transit Survey (NGTS; \citealt{NGTS}) consists of twelve independent 0.2-m telescopes, each equipped with a red-sensitive 2k $\times$ 2k CCD covering eight square degrees. The survey is optimised for detecting short-period planets transiting K-dwarf stars, with the goal of detecting Neptune-sized planets. One of the main drivers for the extremely high-precision photometry achieved by NGTS is the telescope guiding, which uses {\sc Donuts} \citep{donuts} to ensure sub-pixel pointing precision throughout the duration of an observing season. This has allowed the detection of significantly shallower transits than previously achieved from the ground, and the discovery of planets in the Neptunian desert, such as NGTS-4b \citep{NGTS-4}.

In this paper we report the discovery from NGTS of a transiting planet slightly larger than Neptune orbiting the early K-dwarf \starq. In Section~\ref{sec:companion}, we present the host star, and show that it probably has a bound long-period M-dwarf companion. In Section~\ref{sec:obs} we present our observations of the system with both NGTS and other facilities. In Section~\ref{sec:star} we characterise the host star, and in Section~\ref{sec:analysis} we perform a joint analysis of the transit photometry and radial velocities to determine the system parameters. Finally, we discuss \system in the context of the Neptunian desert in Section~\ref{sec:discuss}, and summarise our findings in Section~\ref{sec:conclusions}.

\section{\companion}
\label{sec:companion}

\begin{figure}
\centering
\includegraphics[width=8cm]{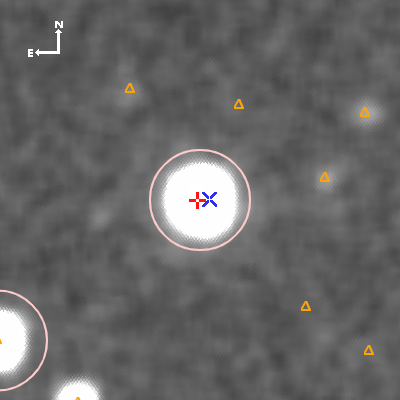} 
\includegraphics[width=8cm]{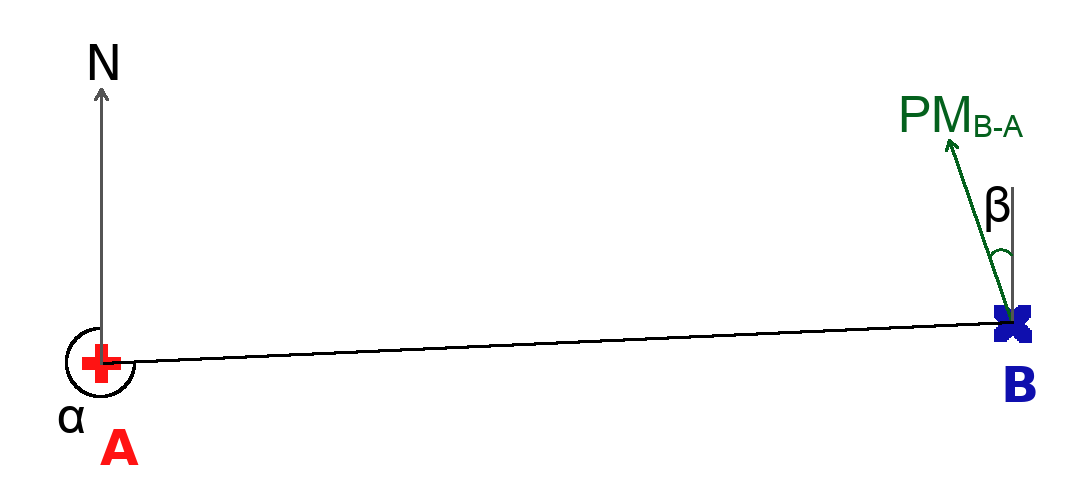}
\caption{Upper panel: NGTS deep stacked image ($2^\prime \times 2^\prime$) centred on \starq. The Gaia positions of \starq and \companion are indicated with a red '+' and a blue 'x', respectively. The pink circle indicates the NGTS photometric aperture, which has a radius of 3 pixels ($15^{\prime\prime}$). Other Gaia sources are marked with orange triangles. North is up, and East to the left.
Lower panel: Sketch of \system, showing the separation of \starq and \companion, which are represented by the same symbols as in the upper panel. \companion lies on a bearing $\alpha = 267.49$ degrees from \starq. The direction of the relative proper motion of \companion with respect to \starq is indicated with the green arrow. The bearing of this proper motion vector is $\beta = 18.8 \pm 9.5$ degrees, and its magnitude is $1.18 \pm 0.20$~mas yr$^{-1}$.}

\label{fig:gaia} 
\end{figure} 

We list in Table~\ref{tab:stellar} the Gaia positions, proper motions and parallaxes \citep{GaiaDR2} for both \starq and \companion, a fainter star which lies just $3.59$ arcseconds away (see Fig.~\ref{fig:gaia}). The parallaxes of the two objects are identical to much less than 1~$\sigma$ significance, and the proper motions in the RA direction are compatible at the 2~$\sigma$ level. We note, however, that in the Declination direction, the proper motions differ from each other with a significance of 5.7~$\sigma$. We conclude, however, that these two stars are co-moving, and likely to be gravitationally bound.

As a test of the statistical significance of finding a companion with similar parallax and proper motions close to our target, we searched the Gaia DR2 catalogue for all stars lying within $5^{\circ}$ of \starq\footnote{ Gaia DR2 reports a total of $4.78 \times 10^5$ sources within $5^{\circ}$ of \starq.} and whose proper motions and parallax match those of \starq to within 6~$\sigma$. This search of an area $2.5\times 10^7$ times larger than the area defined by a radius equal to the separation between target and putative companion returned exactly 100 results, excluding \starq itself. This implies that the chance of finding such an object within $3^{\prime\prime}.6$ is around $4\times 10^{-6}$.

At a distance of $316.7 \pm 4.8$~pc, the projected separation of the two stars implies a physical separation of $1137 \pm 17$~au. If the two objects are gravitationally bound, this sets a lower limit to the binary orbital period of around 40\,000~yr, assuming a circular orbit. For a moderately eccentric orbit (0.6), this lower limit is reduced to around 20\,000~yr, which is the orbital period obtained if we assume that we are observing \companion at apastron. We note that the physical separation implied here is an order of magnitude smaller than the widest binary that can survive for $\sim 10$~Gyr in the Solar neighbourhood \citep{Jiang_Tremaine_2010}.

We also note that at a distance of around $320$~pc, the apparent discrepancy in proper motions in the declination direction ($1.1 \pm 0.2$~mas yr$^{-1}$) corresponds to a physical velocity of around 1.5~\kms, which could be easily explained by the orbital motion of the secondary with respect to the primary. We searched the Digitized Sky Survey for archival imaging, but we found no images with sufficient angular resolution to resolve the two sources. A radial velocity measurement of \companion would provide further evidence of whether or not the two stars are really a binary pair, but \companion is too faint to be observed with CORALIE or HARPS (See Section~\ref{sec:obs-spec}).

The existence of such binary companions to planet host stars is interesting, since even at these very large separations, a binary companion may influence the evolution of the planet. For instance, long-period outer companions can act to maintain a non-zero planetary orbital eccentricity, even for short-period planets whose orbits would otherwise be rapidly circularised through tidal interactions with the host star (e.g. \citealt{Wu_Murray_03}).

\companion is $4.098 \pm 0.004$ magnitudes fainter than \starq in the $G$-band, and $2.17 \pm 0.08$ magnitudes fainter in the $K$-band. The spectral type of \starq is estimated to be K1V, based on the effective temperature derived from spectral analysis, and the tabulation of \citealt{Pecaut_Mamajek}\footnote{\url{http://www.pas.rochester.edu/~emamajek/EEM_dwarf_UBVIJHK_colors_Teff.txt}}.

Based on the absolute $G$-band magnitudes in the aforementioned table, if \starq has a spectral type of K1V, then \companion is probably an M2.5V star. Alternatively, using the (G-K) colour, the best-matching spectral types (according to the aforementioned table) are K1.5V and M3.5V for \starq and \companion, respectively. We adopt spectral types of K1V and M3V for the two stars, and account for the flux contamination of \companion in our light curves (Sec.~\ref{sec:contamination}).

\begin{table*}
\caption{Catalogue information for \starq and \companion.}
\begin{center}
\begin{tabular}{llll} \hline
Positions \& proper motions  & \starq & \companion & Source \\ \hline
RA (J2015.5) & 21h54m04.20s & 21h54m03.89s & Gaia \\
Dec (J2015.5) & $-38^{\circ}~22^\prime~38^{\prime\prime}.71$ & $-38^{\circ}~22^\prime~38^{\prime\prime}.55$  & Gaia \\
pmRA (mas yr$^{-1}$) & $-23.662 \pm 0.042$  & $ -23.282 \pm 0.191$ & Gaia \\
pmDec (mas yr$^{-1}$) & $5.488 \pm 0.034$  & $6.602 \pm 0.192$ & Gaia \\
parallax (mas) & $3.076 \pm 0.034$ & $3.088 \pm 0.176$  & Gaia \\
Projected separation from \starq& -- &$3^{\prime\prime}.59016 \pm 0^{\prime\prime}.00015  $& Gaia \\
Distance from Earth (pc)$^\dagger$ & $316.7 \pm 4.8$ & $316 \pm 18$\\
\hline
Magnitudes& &\\
\hline
$B$ (Johnson)&  $14.093 \pm 0.055$  & -- & APASS \\ 
$V$ (Johnson)&  $13.237 \pm 0.078$ & -- & APASS \\ 
$G$ (Gaia)&  $13.0986 \pm 0.0006$ & $17.197 \pm 0.004$ & Gaia \\ 
$BP$ (Gaia)& $13.5448\pm0.0014$ &  $18.147\pm0.060$ & Gaia \\
$RP$ (Gaia)& $12.5095\pm0.0008$ &  $15.947\pm0.020$ & Gaia \\
$g^\prime$ (Sloan)&  $13.607\pm 0.084$  & -- & APASS \\ 
$r^\prime$ (Sloan)&  $12.989\pm0.054$  & -- & APASS \\ 
$i^\prime$ (Sloan)&  $12.751\pm0.056$ & -- & APASS \\ 
TESS & $12.5638\pm0.006$ & $16.211\pm0.014$ & TESS \\
$J$ &  $11.813\pm 0.029$ & $\geq 12.431$ (2 $\sigma$ limit) & 2MASS\\ $H$ &  $11.386\pm 0.037$ & $\geq 11.859$  (2 $\sigma$ limit) & 2MASS\\ $K$ &  $11.305\pm 0.027$ & $13.474\pm0.070$  & 2MASS\\ 
$W1$ & $11.148 \pm 0.022$ & -- & WISE\\
$W2$ & $11.212 \pm 0.020$ & -- & WISE\\
$W3$ & $11.334 \pm 0.166$ & -- & WISE\\
Spectral type & K1V & M3V & Sec.~\ref{sec:companion}\\
\hline
\multicolumn{3}{l}{Additional identifiers:}\\
\hline
Gaia DR2 & 6585082036193768832 & 6585082036193769088 & \\
2MASS J & 21540423-3822388 & 21540393-3822386 & \\
TIC & 197643976 & 197641898 & \\
\hline
\multicolumn{4}{l}{Gaia: Gaia DR2 \citep{Gaia,GaiaDR2}}\\
\multicolumn{4}{l}{APASS: AAVSO Photometric all-sky survey DR9 \citep{APASS_DR9}}\\
\multicolumn{4}{l}{2MASS: The Two Micron All Sky Survey \citep{2MASS}}\\
\multicolumn{4}{l}{WISE: Wide-field Infrared Survey Explorer \citep{ALLWISE}}\\
\multicolumn{4}{l}{$^\dagger$Calculated from Gaia DR2 parallax, incorporating the systematic offset of \cite{Stassun_Torres_2018}}\\
\end{tabular}
\end{center}
\label{tab:stellar}
\end{table*}

\section{Observations}
\label{sec:obs}

\subsection{NGTS photometry}
\label{sec:obs-ngts}

\starq lies in one of the fields observed by NGTS starting from the commencement of routine science operations in 2016 April. In all, a total of 220\,829 observations were made of this field using a single NGTS telescope, between 2016 April 21 and 2016 December 22. Each image has an exposure time of 10~s, and a typical cadence of 13~s. The data were reduced using the NGTS pipeline \citep{NGTS}, which comprises standard bias subtraction and flat-field correction, followed by aperture photometry based on the CASUtools\footnote{\url{http://casu.ast.cam.ac.uk/surveys-projects/software-release}} software package. Finally, systematic effects in the light curves were removed using the {\sc SysRem} algorithm \citep{Tamuz-etal05}.

The NGTS light curves are searched for period transit-like signals using our implementation of the box-fitting least-squares (BLS) algorithm \citep{Kovacs-etal02}. In the case of \starq, a signal consistent with a transiting exoplanet was detected, with a periodicity of around 3.5~days. After passing several vetting checks designed to eliminate false positives such as eclipsing binaries and blended systems \citep[e.g.][]{Guenther2017}, we initiated follow-up observations to confirm the planetary nature of the system, and to better characterise it.

We also note that this system received a high planetary probability of 0.97 from a neural network trained to distinguish between transiting planetary systems and false positives in NGTS data \citep{NGTS_CNN_1}.

\subsection{Follow-up photometry}
\label{sec:obs-phot}

We used several larger aperture telescopes, as well as multiple NGTS telescopes, to perform follow-up transit observations. The motivation of these observations was to confirm the transit, increase the signal-to-noise of our transit photometry to improve the characterisation of the system, and to improve our knowledge of the orbital ephemeris by extending the observational baseline.

\subsubsection{SAAO 1-m}
We observed a partial transit of \planetq on the night of 2019 October 04, from the South African Astronomical Observatory in Sutherland, Northern Cape using their 1-m telescope. The telescope was equipped with one of the Sutherland High-speed Optical Cameras (SHOC; specifically `SHOC'n'awe'). A description of the instrument can be found in \cite{SHOC}. The observations were conducted in I-band, in focus, with $4\times4$ binning of the CCD, and each of the 960 exposures had a duration of 20~s.

The data were bias and flat field corrected via the standard procedure, using the \textsc{safphot} Python package (Chaushev \& Raynard, in preparation). Differential photometry was also carried out using \textsc{safphot}, by first extracting aperture photometry for both the target and comparison stars using the \textsc{SEP} package \citep{Barbary2016}. The sky background was measured and subtracted using the \textsc{SEP} background map, adopting box size and filter width parameters that minimised background residuals, measured across the frame after masking the stars. Three comparison stars were then utilised to perform differential photometry on the target, using a 3.4 pixel ($0^{\prime\prime}.57$) radius aperture which maximised the signal-to-noise.

\subsubsection{EulerCam}

A partial transit of \planetq was observed on the night of 2019 August 08 with EulerCam on the 1.2-m Euler-Swiss telescope at La Silla Observatory, Chile. A total of 271 observations were made in a filter corresponding to the NGTS bandpass, each with an exposure time of 40~s, giving a cadence of around 52~s. A slight defocus was applied during the observations. Standard data reduction and aperture photometry techniques were applied, using a photometric aperture radius of 19 pixels ($= 4^{\prime\prime}.085$), and 17 comparison stars.

\subsubsection{CHAT}
Another partial transit of \planetq was observed from the Chilean-Hungarian Automated Telescope (CHAT), a 0.7-m robotic telescope installed at Las Campanas Observatory, Chile. The observations were conducted on the night of 2019 November 8, using an $i^{\prime}$ filter. The cadence of the light curve is around 180~s and 73 images were obtained spanning airmass values between 1 and 2. Data was processed with a dedicated pipeline adapted from a set of routines to process photometric data of the LCOGT network \citep[e.g. ][]{hartman:2019, espinoza:2019, jordan:2019}.

\subsubsection{NGTS multi-telescope observations}

In addition to its survey mode, where each telescope observes a different field to maximise sky coverage, NGTS can also be operated in a mode where multiple telescopes observe the same target simultaneously \citep{NGTS_multi}. This mode is used for follow-up of shallow transits detected by the NGTS survey, as well as for TESS targets \citep{ltt9779, TOI-849} and for other exoplanets transiting bright stars \citep{bryant20multicam}.

We observed a partial transit of \planetq on the night of 2019 November 08 (the same event we observed with CHAT), using eight of the NGTS telescopes. We obtained a total of 8790 exposures (around 1100 per camera), each with the usual NGTS exposure time (10~s) and cadence (13~s).

The NGTS follow-up data for \starq were reduced using a custom photometry pipeline. This pipeline uses the \textsc{SEP} Python library for both the source extraction and the aperture photometry \citep{bertin96sextractor, Barbary2016}. We do not apply bias, dark, and flat-field image corrections during the image reduction, as testing showed them to provide no improvement to the photometric precision achieved. We use Gaia \citep{GaiaDR2} to automatically identify comparison stars with a similar colour, brightness and CCD position to \starq  \citep[see][for more details on the photometric pipeline]{bryant20multicam}.

\subsubsection{TESS}

\starq falls in Sector 1 of the TESS survey, which was observed between 2018 July 25 and 2018 August 22, covering seven transits of \planetq. We were able to extract a light curve from the full-frame images (FFIs) of Camera 1, CCD 2, which have a cadence of 30~minutes. To do this, we used a custom photometric aperture selected on the basis of a flux threshold, and to minimise blending from other sources. A $15 \times 15$ pixels area was used for background estimation.

We ran the BLS algorithm on the TESS data to test if the planet could have been detected from the TESS data alone. A peak at around 3.5~d is the highest in the resulting periodogram, with a significance (following \citealt{Cameron-etal06}) of 10.9. This suggests that an exhaustive search of light curves derived from the TESS FFIs could have revealed this system.  However the transit is much less evident in TESS data compared with the NGTS data.  This is primarily due to the fact that NGTS typically produces higher precision photometry for stars with T$>$12.5.

The transit photometry from all the instruments listed above is listed in Table~\ref{tab:phot}, and shown in Fig.~\ref{fig:phot}.

\begin{table}
	\centering
	\caption{Photometry of \starq. The full table is available at the CDS. Only a few lines are shown here for guidance on the format. Note that the fitted offsets between datasets are not applied here, and nor is our correction for contamination from \companion.}
	\label{tab:phot}
	\begin{tabular}{cccc} 
\hline
$\mathrm{BJD_{UTC}}	$		&	Relative		&$\sigma_\mathrm{flux}$  &	Inst. \\
(-2450000)	& flux &  &  \\
		\hline
7562.688132 & 1.014176 & 0.006391 & NGTS \\
7562.689019 & 0.993882 & 0.005278 & NGTS \\
7562.690361 & 0.993867 & 0.005867 & NGTS \\
7562.691707 & 0.997182 & 0.005817 & NGTS \\
7562.693124 & 0.992255 & 0.005179 & NGTS \\
7562.694540 & 1.001568 & 0.002645 & NGTS \\
7562.695955 & 0.990161 & 0.004289 & NGTS \\
7562.697372 & 0.991594 & 0.005725 & NGTS \\
7562.698713 & 0.991647 & 0.005552 & NGTS \\
7562.700052 & 0.992255 & 0.005312 & NGTS \\
		\hline
	\end{tabular}
\end{table}

\subsection{High resolution spectroscopy}
\label{sec:obs-spec}

\subsubsection{CORALIE}
We observed \starq with the CORALIE spectrograph on the Swiss 1.2 m Euler telescope at La Silla Observatories, Chile \citep{CORALIE}, between 3 November and 4 December 2018. CORALIE has a resolving power of $R\sim\numprint{60000}$ and is fed by a 2\arcsec\ on-sky science fibre. 
Radial velocity (RV) measurements were computed for each epoch by cross-correlating with a binary G2 mask \citep{Pepe2002}. The observations showed no significant RV variation within the uncertainties of $\sim 46\, \ms$ and were used to screen \starq for possible scenarios of blended eclipsing binaries mimicking a planetary transit.

\subsubsection{HARPS}
We observed \starq using the HARPS spectrograph \citep{Mayor2003} on ESO's 3.6-m telescope at La Silla Observatories, Chile, under programmes 0103.C-0719 and 0104.C-0588. A total of fifteen measurements were made between 2019 August 03 and 2019 September 20. 

We used the standard HARPS data reduction software to obtain RV measurements of \starq at each epoch. This was done via cross-correlation with a K5 binary mask. Bisector-span, FWHM and other line-profile diagnostics were computed as well. Using an exposure time of 2700~s we obtained typical error bars of $\sim 5\, \ms$. The resulting radial velocities are listed in Table~\ref{tab:rv}, and are plotted in Fig.~\ref{fig:rv}.

The HARPS science fibre has a 1\arcsec\ on-sky projection, and does thus not include \companion. We co-added the fifteen HARPS spectra while weighting each epoch the inverse-variance. Section \ref{sec:spec} details the spectral analysis performed on the stacked spectrum.

\begin{table}
	\centering
	\caption{Radial Velocities for \starq}
	\label{tab:rv}
	\begin{tabular}{ccccc} 
\hline
$\mathrm{BJD_{UTC}}	$		&	RV		&RV err &	FWHM& 	BIS\\
(-2450000)	& (\ms)& (\ms)&(\ms) & (\ms) \\
		\hline
8698.71245 & 30\,272.51 &  4.38  & 6\,147.32 & 8.36 \\
8699.65656 & 30\,294.53 &  5.40  & 6\,139.47 & 5.61 \\ 
8700.69210 & 30\,299.43 &  4.63  & 6\,153.15 & -1.84 \\
8717.70010 & 30\,298.77 &  8.05  & 6\,160.60 & 13.47 \\
8718.60361 & 30\,297.13 &  4.38  & 6\,161.24 & 7.71 \\
8719.73729 & 30\,276.65 &  4.14  & 6\,157.38 & 11.83 \\
8721.65430 & 30\,302.65 &  4.81  & 6\,158.79 & 2.14 \\
8722.60994 & 30\,277.93 &  4.79  & 6\,140.92 & 20.31 \\
8723.58722 & 30\,276.65 &  4.62  & 6\,131.25 & 10.87 \\
8724.53490 & 30\,295.00 &  7.17  & 6\,141.78 & 12.75 \\
8724.80470 & 30\,299.44 &  5.63  & 6\,120.43 & 20.12 \\
8725.56246 & 30\,292.16 &  5.89  & 6\,124.07 & 4.17 \\
8725.74721 & 30\,287.79 &  7.45  & 6\,134.53 & 43.93 \\
8730.60832 & 30\,282.33 &  7.34  & 6\,131.68 & 21.46 \\
8746.58959 & 30\,287.98 &  7.77  & 6\,119.95 & 8.34 \\
		\hline
	\end{tabular}
\end{table}

\section{Stellar characterisation}
\label{sec:star}

\subsection{Spectral analysis} \label{sec:spec}

The HARPS spectra were co-added, and the resulting spectrum (S/N $\approx 49$) was analysed with the synthesis method (without the use of wavelets) outlined in \cite{Gill18}. The resulting parameters are $\teff = 5200 \pm 85$~K, $\feh = 0.12 \pm 0.08$, $\logg = 4.2\pm0.1$ (cgs), and $\vsini = 1.2\pm0.5$~\kms.

\subsection{SED fit with {\tt ARIADNE}}
\label{sec:ariadne}

\begin{figure} 
\includegraphics[width=8.8cm, angle=0] {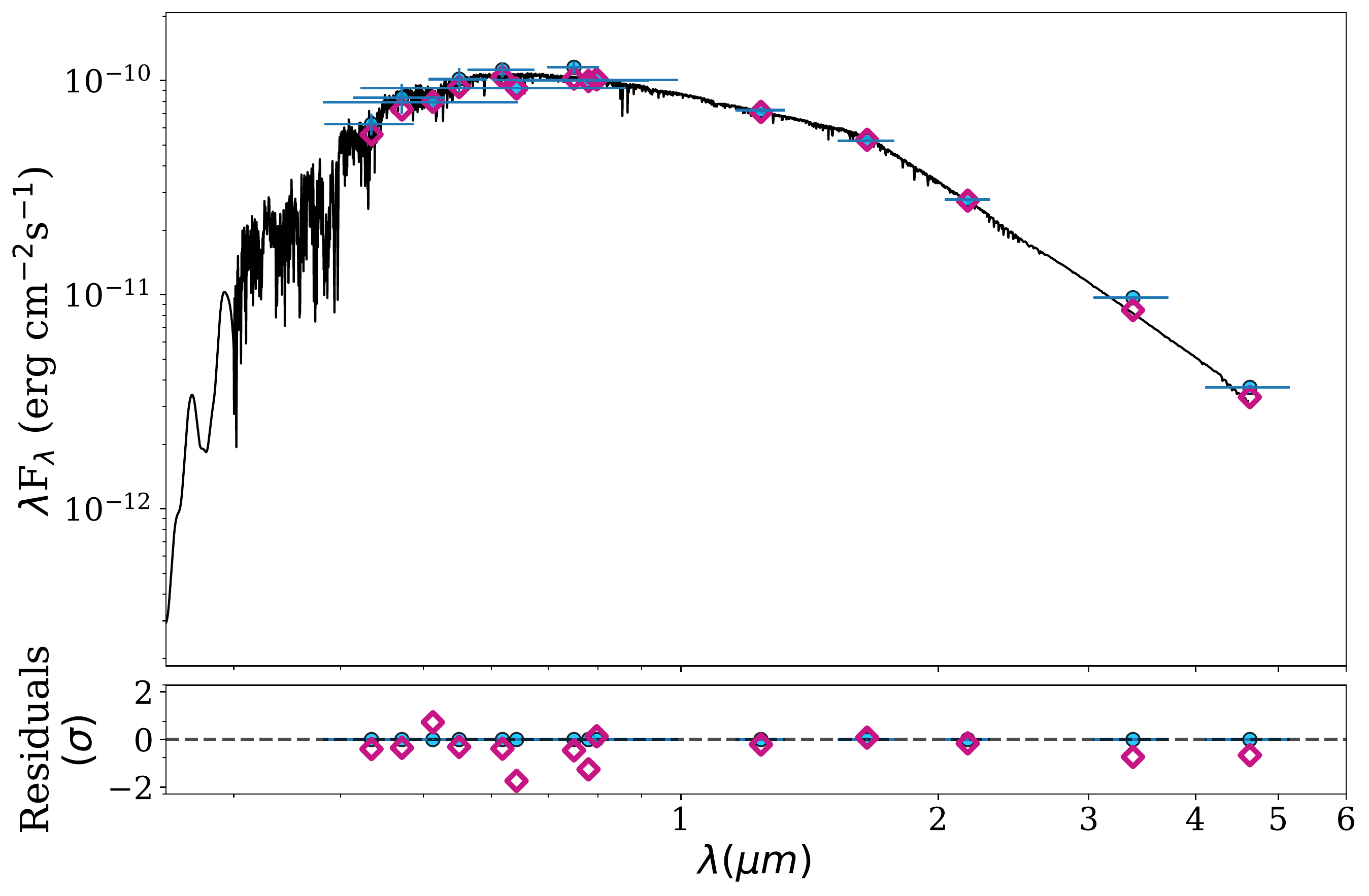} 
\caption{Spectral energy distribution (SED) of \starq. Catalogue photometric measurements (Table~\ref{tab:stellar}) are shown as blue circles, with the horizontal error bars indicating the extent of each bandpass. A model SED from \cite{Castelli2004} is shown as a solid black line, with magenta squares indicating the model flux integrated over each bandpass.
}
\label{fig:sed} 
\end{figure}

We fit the spectral energy distribution (SED) of \starq using \texttt{ARIADNE}, a python tool written to fit
catalogue photometry to different atmospheric model grids (Fig.~\ref{fig:sed}). It hosts model grids for
\texttt{Phoenix v2} \citep{Husser2013}, \texttt{BT-Settl}, \texttt{BT-Cond}, \texttt{BT-NextGen}
\citep{Allard2012, Hauschildt99}, \cite{Castelli2004}, and \cite{Kurucz1993} convolved with filter response
functions: $UBVRI$; 2MASS $JHK_{rm s}$; SDSS \textit{ugriz}; WISE $W1$ and $W2$; Gaia $G$, $RP$, and $BP$,
Pan-STARRS $griwyz$' Str\"omgren uvby; GALEX NUV and FUV; {\it TESS}; {\it Kepler}; and NGTS. Each synthetic
SED is modelled by interpolating grids in \teff-\logg-\feh space with the remaining parameters being
distance, radius, extinction in the $V$ band, and an individual excess noise for each photometry in order to
account for underestimated uncertainties. The priors for \teff, \logg and \feh were taken from the spectral
analysis determined values, for radius and distance we took the Gaia DR2 reported value, noting that we
inflated the radius reported error to account for modelling errors and we applied the
\citealt{Stassun_Torres_2018} correction to the parallax, we limited the $A_{V}$ to the maximum
line-of-sight taken from the re-calibrated SFD galactic dust map (\citealt{Schlegel1998, Schlafly2011}), and
finally, each excess noise parameter has a zero mean Normal distribution as their priors with a variance
equal to five times the size of the reported uncertainty. \texttt{ARIADNE} uses \texttt{dynesty}'s nested
sampler for parameter estimation and calculating the Bayesian evidence for each model \citep{Speagle2019}. The final step in \texttt{ARIADNE}'s algorithm is calculating the weighted average of each parameter using the relative probabilities of each models as weights. For a detailed explanation of the fitting procedure, accuracy and precision of \texttt{ARIADNE} the reader is referred to \citet{Vines2020}. The resulting stellar parameters are listed in Table~\ref{tab:stellar_params}.

\begin{table}
\caption{Stellar parameters}
\begin{tabular}{lll} \hline
Parameter  & Value & Source\\ \hline
\teff / K     &   $5187\pm11$ & {\tt ARIADNE}\\
\logg (cgs)     &  $4.20\pm0.04$ & {\tt ARIADNE}\\
\vsini / \kms  &   $1.2\pm0.5$ & HARPS spectrum \\
\feh & $0.10\pm0.03$ & {\tt ARIADNE}\\
\rstar / \rsol & $0.842 \pm 0.006$ & {\tt ARIADNE}\\
\mstar / \msol &$0.898\pm0.035$ & {\tt ARIADNE}\\
Age / Gyr & $5.9^{+3.0}_{-3.4}$ & {\tt ARIADNE}\\
\hline
\end{tabular}
\label{tab:stellar_params}
\end{table}

\subsection{TIC}

The TESS Input Catalog \citep{TIC} lists the following parameters for \starq, based on the Gaia DR2 \citep{GaiaDR2} observations: $\rstar = 0.856 \pm 0.049$~\rsol, $\teff = 5222 \pm 128$~K. We note that our adopted values are in good agreement with these TIC values.

\begin{figure} 
\includegraphics[width=\textwidth, angle=270] {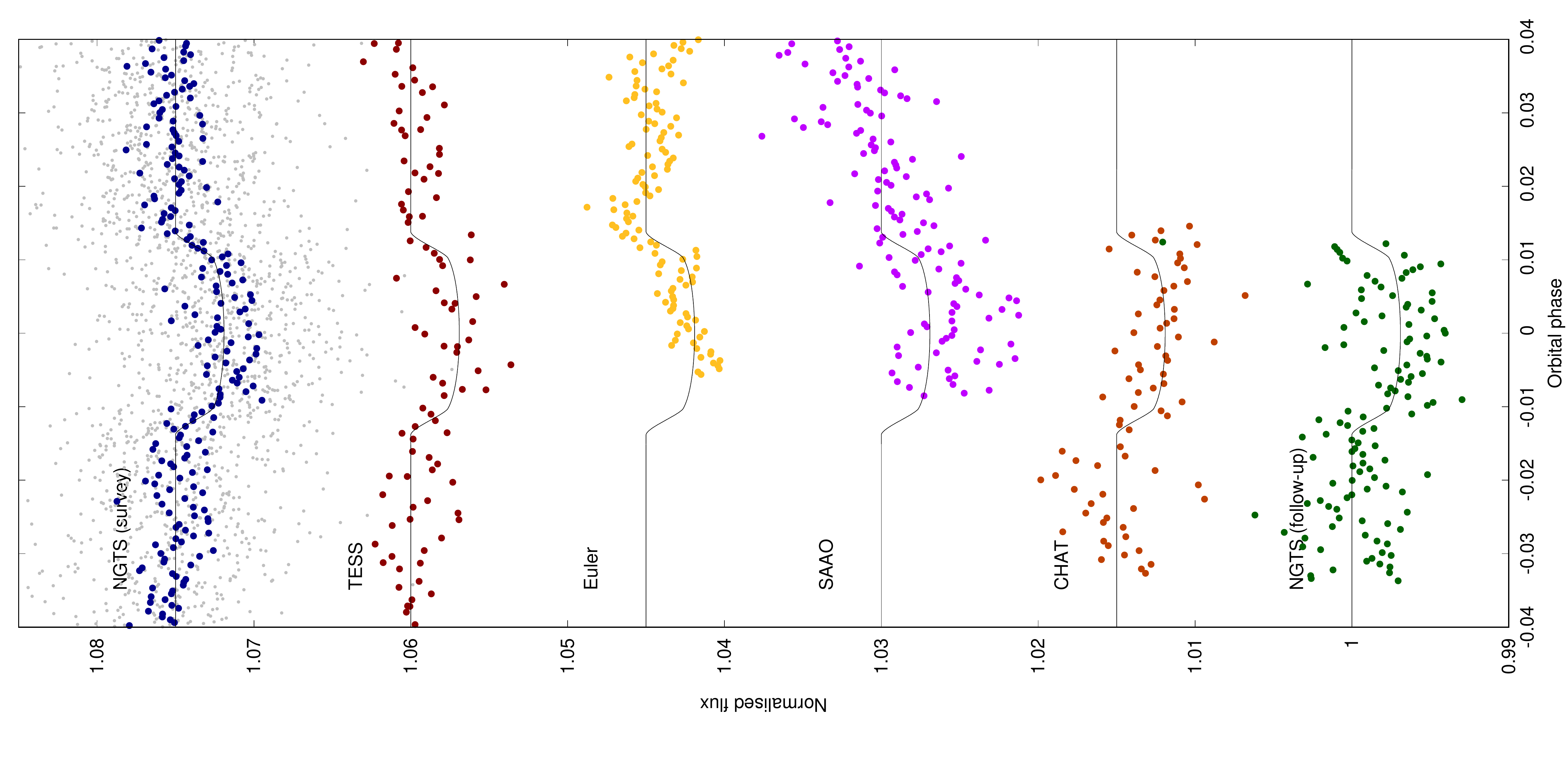} 
\caption{Transit photometry of \starq from several instruments, offset vertically for clarity. From top-to-bottom: NGTS survey photometry (binned to two minutes in grey, binned to two minutes equivalent in phase in blue); TESS (unbinned); Euler (binned to two minutes); SAAO (binned to two minutes); CHAT (unbinned); single night NGTS multi-telescope observations (binned to two minutes). Our best-fitting model is shown as a solid black line.
}
\label{fig:phot} 
\end{figure} 

\begin{figure} 
\includegraphics[width=12cm, angle=270]{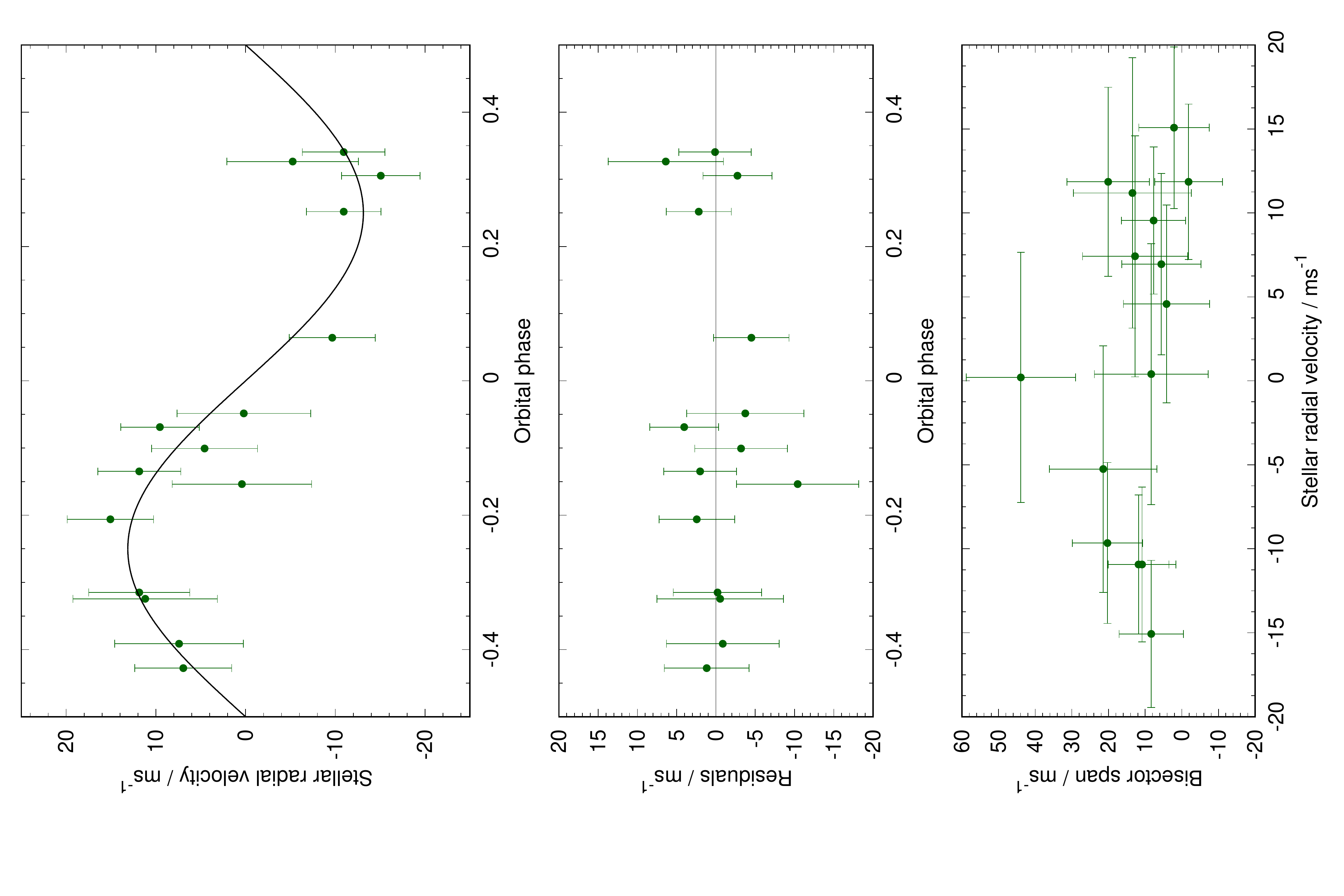} 
\caption{Radial velocities of \starq from HARPS. Upper panel: RV as a function of orbital phase, with best-fitting model shown as a solid black line. Middle panel: residuals to the best-fitting model. Lower panel: bisector span of the stellar lines against radial velocity. The uncertainties on the bisector spans are taken to be twice the uncertainty on the radial velocities.
}
\label{fig:rv} 
\end{figure} 

\subsection{Stellar rotation}

We searched the NGTS survey light curve for periodic signals, using both a Lomb-Scargle periodogram and a generalised autocorrelation (G-ACF) method \citep{Blanco1}. Both methods revealed a signal with a period of approximately 10~d to be the most prominent, although given the small amplitude of the putative signal ($\leq$~1~mmag), this is not a robust detection. 

From the \vsini derived from the HARPS spectra ($1.2\pm0.5$~\kms) and the stellar radius, we calculate an upper limit to the stellar rotation period, $\prot \leq 35.5 \pm 14.8$\,d. This value corresponds to the case where the stellar spin and planet rotation axes are aligned.

It is unclear if the 10-d signal is related to rotation, since it would imply a faster-rotating star than suggested by the \vsini, and would be indicative of a star significantly younger (c. 700~Myr; \citealt{rebull17,Douglas17}) than our SED analysis (Sec.~\ref{sec:ariadne}) implies. At an age less than 1~Gyr, K0 stars typically exhibit photometric rotational modulation of at least a few mmag, which would be readily detectable in the NGTS data.

\section{Joint fit to photometry and RVs}
\label{sec:analysis}

We performed a joint fit to the photometry and radial velocities using the Transit and Light Curve Modeller (\tlcm; \citealt{tlcm}). In order to reduce the large number of data points in the NGTS survey photometry, and thus increase modelling speed, we modelled only those nights of NGTS data where a transit occurred, and we binned the remaining light curve to a cadence of 120~s. This resulted in a total of 2404 NGTS survey data points. Similarly, the NGTS multi-telescope data were binned to a single light curve with a cadence of 120~s. We excluded TESS data taken far from transit, only modelling that with an orbital phase between -0.1 and 0.1. The CHAT, SAAO, and Euler light curves were modelled unbinned and in their entirety.

\subsection{Contaminating flux from \companion}
\label{sec:contamination}

In all photometric data sets, except that from the SAAO 1-m, \companion (Sec.~\ref{sec:companion}) is well within the photometric aperture used to extract the target flux from the images. To account for this contamination, we calculated the flux ratio of the two components of \system, and corrected the affected datasets accordingly. We find that \companion contributes around two per cent of the observed flux, and that in the absence of any correction, our value of $R_p/R_*$ is biased by around $2\,\sigma$.

\subsection{Stellar limb darkening}

The signal-to-noise ratio of our photometry is not sufficiently high to precisely constrain the limb-darkening coefficients in each bandpass in which we observed \starq. The bandpasses used do not span a particularly large range of wavelengths; rather they are fairly similar to each other. We calculated theoretically expected quadratic limb-darkening coefficients for each band using the Limb Darkening Toolkit ({\sc LDTk}) of \cite{LDTk}, which relies on the stellar models of \cite{Husser2013}. The difference between the coefficients for the different bandpasses is insignificant, much smaller than the uncertainties on the coefficients we measure for each light curve individually. Further, we find that our derived system parameters are insensitive to our choice of limb-darkening parameters, within this range. We therefore opt to fit for a single pair of limb-darkening coefficients, common to all observed bandpasses. In TLCM, this is done with the parameters $u_+ = u_a + u_b$ and $u_- = u_a - u_b$, where $u_a$ and $u_b$ are the commonly used quadratic coefficients.

\subsection{Orbital eccentricity}
\label{sec:ecc}
We also tried fitting for an eccentric orbit, via the additional parameters $\sqrt{e}\sin(\omega)$ and $\sqrt{e}\cos(\omega)$, where $e$ is the orbital eccentricity, and $\omega$ the argument of periastron. Our best-fitting values are $\sqrt{e}\sin(\omega) = 0.07 \pm 0.27$ and $\sqrt{e}\cos(\omega) = -0.05 \pm 0.20$, resulting in $e= 0.0074 \pm 0.0428$ and $\omega = -54 \pm 151$ degrees. The three-sigma upper limit to the eccentricity is 0.11. The very modest improvement in $\chi^2$ of less than 0.4 does not compensate for the much larger Bayesian penalty on having two additional model parameters. This results in a significantly lower BIC (Bayesian Information Criterion) for the circular orbital model ($\Delta\,\mathrm{BIC_{circ-ecc}} = -5.05$). We therefore reject the eccentric model, and adopt $e=0$.

\subsection{Additional trend in RVs}

We also tried fitting for a radial acceleration, $\dot{\gamma}$, of \starq. The presence of such a trend in the RVs is usually indicative of an additional body in the system (e.g. \citealt{K299}). We obtained a best-fitting value $\dot{\gamma} = -9 \substack{+21 \\ -17}\, \mbox{m\,s$^{-1}$\,yr$^{-1}$}\xspace$, suggesting that no significant radial acceleration is detected. This is confirmed by a BIC analysis, where the simpler model with no acceleration has a smaller BIC value (by 1.2), and is hence preferred.

\subsection{Bisector spans}

We observe no evidence for a correlation between either time and bisector span or bisector span and radial velocity (Fig.~\ref{fig:rv}). A correlation can indicate that the apparent planetary signal is caused by a blended eclipsing binary system, or by stellar activity \citep{Queloz01}. Furthermore, we see no correlation between radial velocity and the FWHM of the line profile.

Our joint fit to the photometry and RVs is shown in Figs.~\ref{fig:phot} and \ref{fig:rv}, respectively. The resulting system parameters are listed in Table~\ref{tab:param}.

\begin{table*}[!th]
\begin{center}
\caption{Parameters from light curve and RV data analysis.\label{tab:param}}
\begin{tabular}{lcc}
\hline
\hline
\noalign{\smallskip}
Parameter & value & unit\\
\noalign{\smallskip}
\hline
\noalign{\smallskip}
{\sc TLCM} fitted parameters: &&\\
\hline
\noalign{\smallskip}
Orbital period $P_\mathrm{orb}$  & 		\PERIOD	 & d\\
Transit epoch $T_0$ 			 & 		\EPOCH	 &BJD$_\mathrm{TDB}-2450000$\\
Scaled semi-major axis $a/R_*$ 	 &    	\ARSTAR	&\\
Radius ratio $R_p/R_*$ 	 &  \RPRS &\\
Transit impact parameter $b$     &     \IMPACT & \\
Limb-darkening coefficient $u_+$ &   \UPLUS &\\
Limb-darkening coefficient $u_-$ &   \UMINUS  &\\
Radial velocity semi amplitude $K$ &  \KAmp  & \ms \\
Systemic radial velocity $\gamma$ & \GAMMA &  \kms  \\
\noalign{\smallskip}
Derived parameters:&&\\
\hline
\noalign{\smallskip}
Eccentricity $e$ & 0.0 & (fixed, see Sec.~\ref{sec:ecc}) \\
Semi-major axis $a$&    \SMA	  & au\\
Orbital inclination angle $i$	 &    \INCLINATION    &$\deg$\\
Transit Duration $T_{14}$   	 & 		\DURATION			 & h\\
Planet mass $M_\mathrm{p}$   			& \SecondaryMass   &$M_\mathrm{Jup}$\\
Planet radius $R_\mathrm{p}$ 			& \SecondaryRadius  &$R_\mathrm{Jup}$ \\
Planet mean density          & \SecondaryDensity & kg\,m$^{-3}$\\
Planet surface gravity \loggp & \SecondaryLogg   &(cgs)\\
Planet equilibrium temperature$^\dagger$ $T_\mathrm{p,A=0}$ & \SecondaryTeff  &K\\
\noalign{\smallskip}
\hline
$^{\dagger}$assuming zero albedo, and isotropic heat redistribution.\\
\end{tabular}
\end{center}
\end{table*}

\section{Discussion}
\label{sec:discuss}

\subsection{\planetq in the Neptunian desert}

\begin{figure*}
\includegraphics[width=\textwidth]{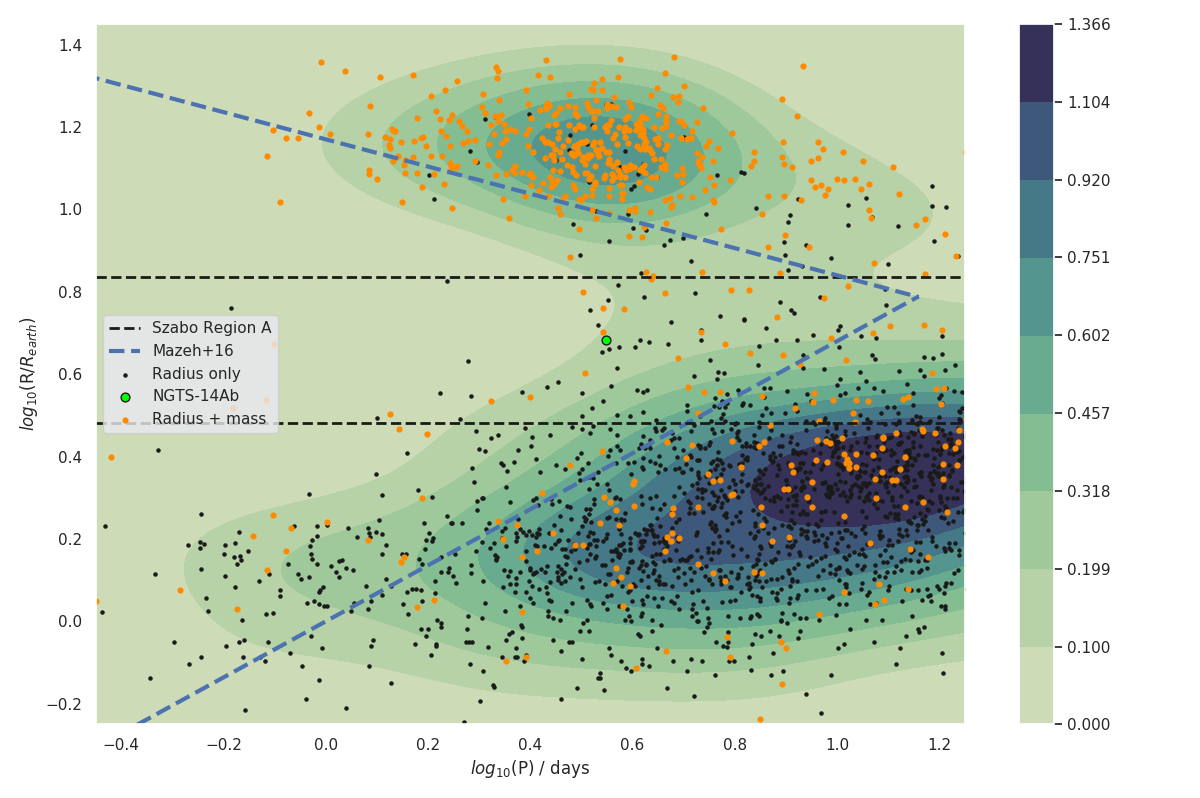} 
\caption{\planetq (light green point) within the Neptunian desert, the boundary of which (as defined by \citealt{Mazeh16}) is indicated with blue dashed lines. The black dashed lines indicate the region of interest identified by \cite{Szabo19}. The parameters of other planets come from the NASA Exoplanet Archive, and the colour scale indicates the number of planets per grid element. The plot comprises $200\times 200$ elements, uniformly spaced in $\log_{10}P$ and $\log_{10}R$.
}
\label{fig:nep} 
\end{figure*} 

\planetq joins a growing number of planets that reside in the Neptunian desert, a sparsely populated region of parameter space between the hot Jupiters and the super Earths in the radius -- period plane (Fig.~\ref{fig:nep}). \planetq lies within both the desert as defined by \cite{Mazeh16}, and the smaller region identified by \cite{Szabo19}, who along with \cite{ngts5} have recently investigated the boundaries of the Neptunian desert as a function of host star spectral type.

\cite{Owen+Lai_2018} explain the lower boundary of the desert as a consequence of photoevaporation --  a process that is more efficient at reducing the size of close-in planets. Further, they suggest that the upper boundary of the desert is a result of the fact that only the highest mass planets can be tidally circularised in the closest orbits, after high-eccentricity migration. \planetq lies among the lower boundaries of the desert computed by \cite{Owen+Lai_2018}, which are dependent upon metallicity and core mass. Comparing \planetq to their simulations (which ran for 5~Gy, similar to the age of the system) suggests that \planetq has a core mass of close to 10~\me, and that any additional envelope accreted by the planet during its formation would have by now been lost to photoevaporation.

\begin{figure} 
\includegraphics[width=10cm]{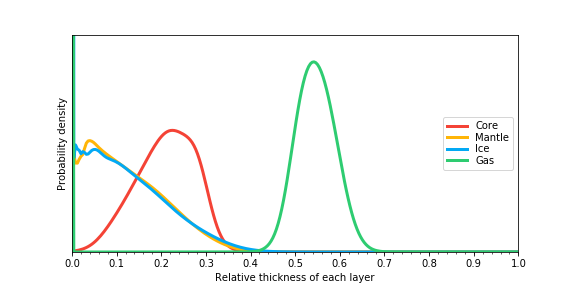} 
\caption{Possible internal structure for \planetq, the result of the neural network model of \cite{Baumeister20}. Shown are the predictions of the relative interior layer thicknesses, using our derived planetary mass and radius as inputs. The colored lines show the combined Gaussian mixture prediction of the model, with the area under each curve normalised to unity.
}
\label{fig:interior} 
\end{figure} 

\begin{figure} 
\includegraphics[width=8cm]{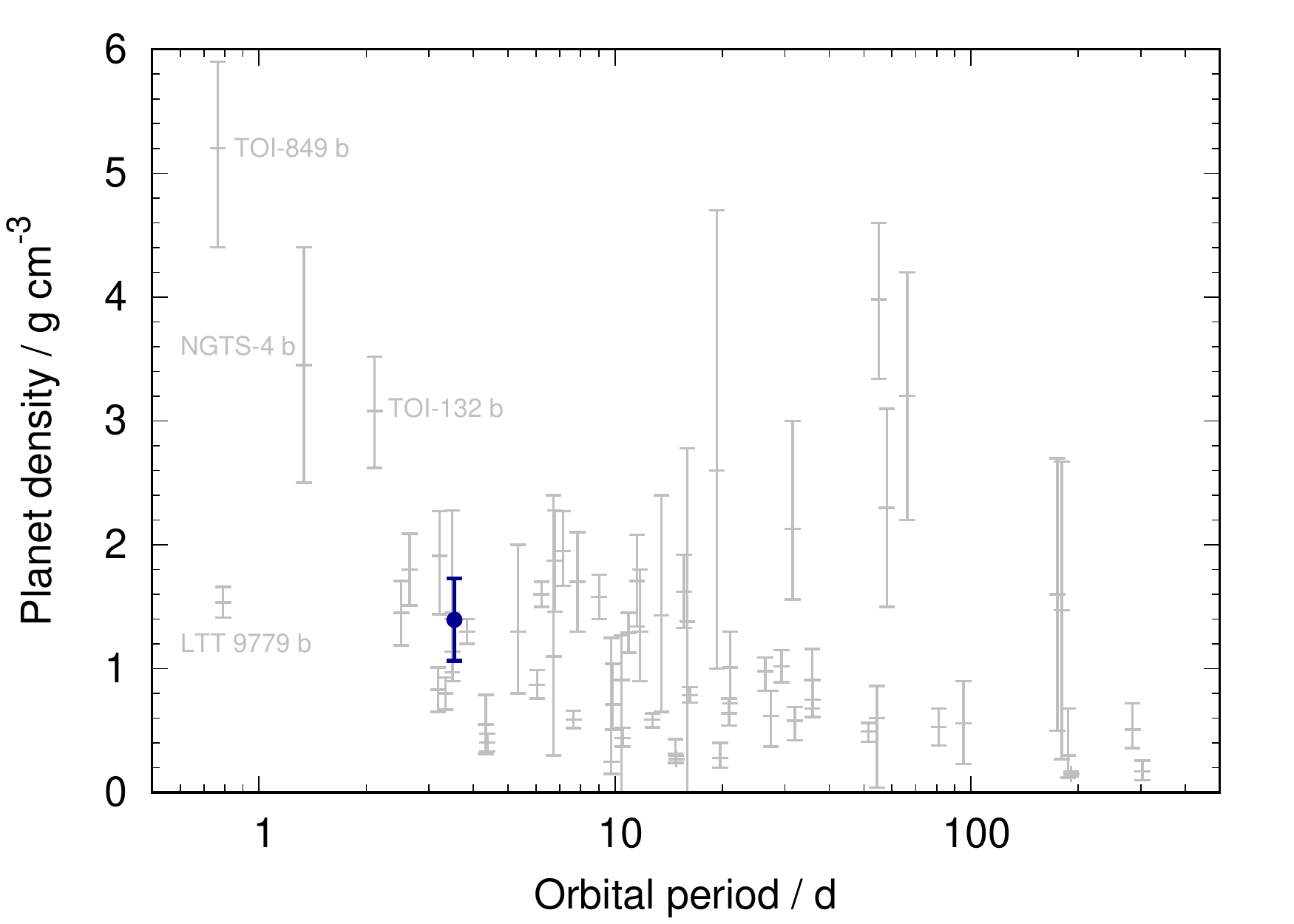}
\includegraphics[width=8cm]{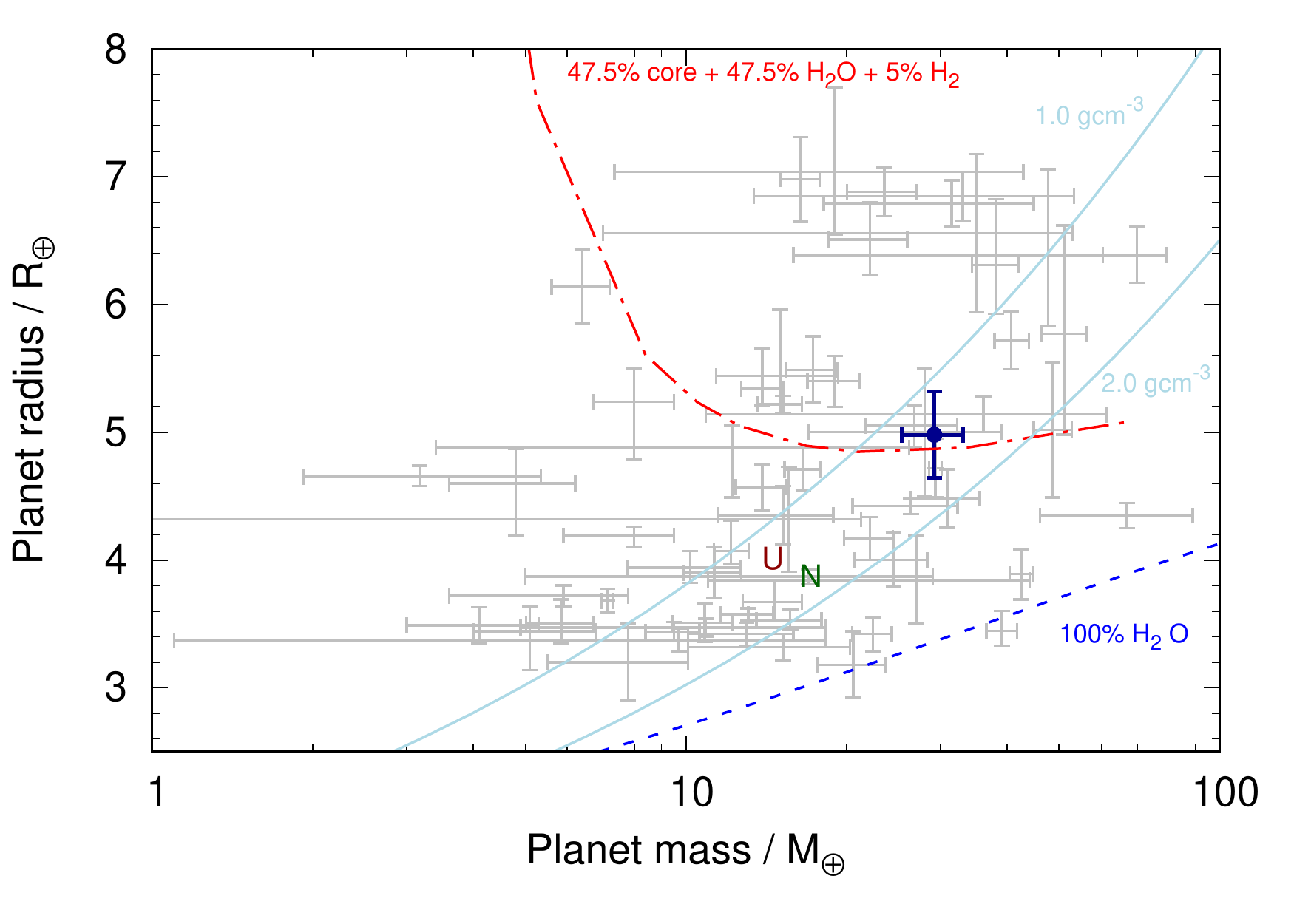}
\caption{\planetq compared to other planets within the Neptunian desert, as defined by \cite{Szabo19} (their `Region A' -- the area between the dashed black lines in Fig.~\ref{fig:nep}). Planetary data is taken from the NASA Exoplanet Archive. \planetq is denoted by a dark blue point in both panels. Upper panel: Planet density versus orbital period. Several planets discussed in the text are labelled with their names. Lower panel: planet radius versus mass. The Solar System ice giants are indicated with their initial letters, and lines of constant density (1.0\,g\,cm$^{-3}$ and 2.0\,g\,cm$^{-3}$) are shown with solid pale blue lines. Two models from \cite{Zeng19} corresponding to compositions of pure water (dashed blue line) and an Earth-like rocky core with an equal mass of water, and five per cent hydrogen (dash-dotted red line) are also shown.}
\label{fig:mass_radius} 
\end{figure} 

We use the neural network model of planetary interiors developed by \cite{Baumeister20} to predict the internal structure of \planetq (Fig.~\ref{fig:interior}), using the planetary mass and radius (Table~\ref{tab:param}) as model inputs. Although internal structure is notoriously degenerate with mass and radius, this model suggests that \planetq has a significant gaseous envelope. The models of \cite{Howe14} suggest that an H$_2$-He envelope constituting approximately half of the planet's radius would contribute five to ten per cent of the planet's mass, for a core mass of 10~\me. In other words, there is broad compatibility between the models of \cite{Baumeister20} (Fig.~\ref{fig:interior}) and the model of \cite{Zeng19} with a five per cent (by mass) hydrogen-rich envelope (Fig.~\ref{fig:mass_radius}, red curve).

In Fig.~\ref{fig:mass_radius}, we compare \planetq to the other planets that inhabit `Region A' of \cite{Szabo19}, i.e. those planets having radii between 0.28 and 0.63 \rjup. Planets orbiting closer than \planetq does, such as NGTS-4\,b \citep{NGTS-4} and TOI-132\,b \citep{TOI-132}, tend to be denser because they have undergone evaporation, and lost much or all of their primary, hydrogen-dominated atmospheres. The most extreme example of this is the recently-discovered TOI-849\,b \citep{TOI-849}, thought to be the remnant core of a giant planet. LTT\,9779\,b \citep{ltt9779} on the other hand, remains an intriguing exception -- how it has apparently retained its hydrogen-rich envelope is a mystery. \planetq appears to just far enough from its star that it is able to maintain a significant atmosphere, resulting in a density similar to those of Uranus and Neptune.

\subsection{Stellar companion}
Motivated by the apparent stellar companion to \starq (Section~\ref{sec:companion}), and \planetq's position in the Neptunian desert (as defined by \citealt{Mazeh16}), we conducted a search for stellar companions to other planet host stars in and around the desert. The presence of a long-period companion may influence the evolution of a planetary system, for instance by driving high-eccentricity migration \citep{Wu_Murray_03}.

We used the Gaia DR2 catalogue \cite{GaiaDR2} to search for companions to stars that are listed in the NASA Exoplanet Archive\footnote{\url{https://exoplanetarchive.ipac.caltech.edu/}, accessed 2020 June 17.} as hosting planets with orbital periods less than ten days, and masses measured to a precision of less than 50 per cent. We then sub-divided the sample into planets lying above, in, or below the Neptunian desert (as defined in the mass-period plane in \citealt{Mazeh16}). This resulted in 73 systems in the desert, 339 systems above the desert, and 55 systems (70 planets) below the desert.

\begin{table*}
\centering
\caption{Potential companions to planetary host stars, based on analysis of Gaia DR2 data. `Plx \& PM matches' refers to potential long-period companions identified by their matching parallaxes and proper motions. `RUWE > 1.4' refers to stars whose astrometric fits exhibit excess noise, potentially indicating the presence of a binary companion.}
\begin{tabular}{llc|cc|cc} \hline
&Subset  & No. of & \multicolumn{2}{c}{Plx \& PM matches} |& \multicolumn{2}{c}{RUWE > 1.4}\\
&& systems & No. & \% & No. & \% \\ \hline
Exoplanet host stars &&&&&\\
&Desert & 73 & 10 & 13.7  & 4 & 5.5\\
&Above desert & 339 & 62 & 18.3 & 18 & 5.3\\
&Below desert & 55 & 3 & 5.5 & 0 & 0.0 \\
&Above + below & 394 & 65 & 16.5 & 18 & 4.6 \\
&Sum of all host stars &467 & 75 & 15.6 & 22 & 4.6 \\
\hline
Analogous non-host stars &&&&&\\
&Desert & 73 & 1 & 1.4  & 11 & 15.1\\
&Above desert & 339 & 27 & 8.0 & 49 & 14.5\\
&Below desert & 55 & 5 & 9.1 & 14 & 25.5 \\
&Sum of all non-host stars &467 & 33 & 7.1 & 74 & 15.8 \\
\hline

\end{tabular}
\label{tab:gaia_companions}
\end{table*}

We then queried Gaia DR2, using a cone search centred on each object, conducted using the astroquery package \citep{astroquery}. The radius of the search was chosen individually for each object, so that it corresponds to a separation of 10\,000~au. We then applied the same technique as used in Section~\ref{sec:companion} to search for nearby objects whose  parallaxes and proper motions match the planet host star within $n\sigma$. For $n=3$, we identified possible companions to 62 host stars (18.3 per cent) from above the desert, ten (13.7 per cent) in the desert, and three (5.5 per cent) below it. Combining the results for above and below the desert, we find potential companions to 16.5 per cent of host stars outside the desert, which is statistically indistinguishable from the proportion inside the desert. These results are summarised in Table~\ref{tab:gaia_companions}.

To extend our binarity analysis to companions too close to be resolved as individual objects in Gaia DR2, we used the renormalized unit weight error (RUWE) to the Gaia DR2 astrometric solutions \citep{RUWE}. A previous analysis conducted using the RUWE statistic has suggested that there is tentative evidence for increased binarity among hot Jupiter host stars \citep{Belokurov20}. 

We adopt the threshold of RUWE = 1.4 suggested by the Gaia team for distinguishing between well-behaved solutions and those where an unseen additional body may impact the astrometry\footnote{Technical note GAIA-C3-TN-LU-LL-124-01 \url{http://www.rssd.esa.int/doc_fetch.php?id=3757412}}. In Table~\ref{tab:gaia_companions}, we report the number of systems for which the RUWE statistic is greater than 1.4. We fail to find a significant difference between the RUWE distributions for objects inside and outside the Neptunian desert. There is, however, a hint of fewer companions below the desert, with no large RUWE values in this population.

Finally, we attempted to compare the three populations of exoplanet-hosting stars with their non-exoplanet host counterparts. To do this, for each exoplanet host star, we searched Gaia DR2 for other nearby stars with similar properties. Specifically, we required that the $G$-band magnitude be within 0.25, the $G_\mathrm{BP} - G_\mathrm{RP}$ colour within 15 per cent, and the parallax within a factor 2 (to exclude giant stars). Where multiple stars were found to match these criteria, we selected one at random. We then performed the same analysis on these non-exoplanet host stars as on the hosts; the results are reported in the lower part of Table~\ref{tab:gaia_companions}. 

We see fewer potential long-period companions with matching parallax and proper motions amongst the non-host stars, suggesting that the presence of a long-period companion increases the probability of a star hosting an exoplanet in or above the desert. Non-well behaved (RUWE > 1.4) astrometric solutions are more likely to occur around stars not known to host an exoplanet, perhaps indicating that short-period binaries are more common in this population. We note, however, that selection effects may play a role here, in that short-period binarity of an exoplanet candidate host star may complicate both photometric and spectroscopic follow-up, and hence the confirmation of exoplanets orbiting stars with close binary companions.

Although this binarity analysis is inconclusive, future Gaia data releases will include the identification of further gravitationally bound companions to exoplanet host stars. This may allow any trends with planet parameters (such as membership of the Neptunian desert) to be elucidated.

\section{Conclusions}
\label{sec:conclusions}

We have presented the discovery of the NGTS-14 system, which consists of an early K-type dwarf star, \starq, with a long-period companion, \companion, which is likely to be a mid-M dwarf. The primary star is orbited by a transiting exoplanet, \planetq, which is slightly larger and more massive than Neptune. The planet's orbital period of just over 3.5 days places it in the Neptunian desert, a thus far sparsely-populated region of parameter space. \planetq has a density which suggests that it has retained some of its primordial atmosphere.

\begin{acknowledgements}
This work uses data collected under the NGTS project at the ESO La Silla Paranal Observatory. The NGTS facility is operated by the consortium institutes with support from the UK Science and Technology Facilities Council (STFC) under projects ST/M001962/1 and ST/S002642/1. 
This research made use of Astropy,\footnote{http://www.astropy.org} a community-developed core Python package for Astronomy \citep{astropy1, astropy2}.

This work has made use of data from the European Space Agency (ESA) mission
{\it Gaia} (\url{https://www.cosmos.esa.int/gaia}), processed by the {\it Gaia}
Data Processing and Analysis Consortium (DPAC,
\url{https://www.cosmos.esa.int/web/gaia/dpac/consortium}). Funding for the DPAC
has been provided by national institutions, in particular the institutions
participating in the {\it Gaia} Multilateral Agreement.

This research has made use of the NASA Exoplanet Archive, which is operated by the California Institute of Technology, under contract with the National Aeronautics and Space Administration under the Exoplanet Exploration Program.

This publication makes use of data products from the Wide-field Infrared Survey Explorer, which is a joint project of the University of California, Los Angeles, and the Jet Propulsion Laboratory/California Institute of Technology, funded by the National Aeronautics and Space Administration. 

This study is based on observations collected at the European Southern Observatory under ESO programmes 0103.C-0719 and 0104.C-0588.\\ 
We  thank  the  Swiss  National  Science  Foundation  (SNSF) and the Geneva University for their continuous support to our planet search programs. This work has been in particular carried out in the frame of the National Centre for Competence in Research {\it PlanetS} supported by the Swiss National Science Foundation (SNSF). \\ 
This publication makes use of The Data \& Analysis Center for Exoplanets (DACE), which is a facility based at the University of Geneva (CH) dedicated to extrasolar planets data visualisation, exchange and analysis. DACE is a platform of the Swiss National Centre of Competence in Research (NCCR) PlanetS, federating the Swiss expertise in Exoplanet research. The DACE platform is available at \url{https://dace.unige.ch}. \\ 

MNG acknowledges support from MIT's Kavli Institute as a Juan Carlos Torres Fellow. Contributions by authors from the University of Warwick were supported by STFC consolidated grants ST/P000495/1 and ST/T000406/1. DJA acknowledges support from the STFC via an Ernest Rutherford Fellowship (ST/R00384X/1).
EG gratefully acknowledges support from the David and Claudia Harding Foundation in the form of a Winton Exoplanet Fellowship.
Ph.E., A.C., and H.R. acknowledge the support of the DFG priority program SPP 1992 “Exploring the Diversity of Extrasolar Planets” (RA 714/13-1).
R.B.\ acknowledges support from FONDECYT Post-doctoral Fellowship Project 3180246, and from the Millennium Institute of Astrophysics (MAS).\\

Finally, we acknowledge our anonymous referee, whose comments helped to improve the quality of this manuscript.

\end{acknowledgements}

\bibliographystyle{aa}
\bibliography{refs2}

\end{document}